\documentclass[prl,twocolumn,letterpaper]{revtex4}
\usepackage{graphicx}
\usepackage{amsmath}
\usepackage{amssymb}

\begin{document}

\title{
\LARGE A dynamical systems approach to \\  actin-based motility in 
{\it Listeria monocytogenes}}
\author{
Scott Hotton} 
\affiliation{Department of Organismic and Evolutionary Biology, Harvard 
University, Cambridge MA 02138}

\begin{abstract}
    A simple kinematic model for the trajectories of {\it Listeria 
monocytogenes} is generalized to a dynamical system rich enough to exhibit the 
resonant Hopf bifurcation structure of excitable media and simple enough to be
studied geometrically.  It is shown how {\it L. monocytogenes} trajectories 
and meandering spiral waves are organized by the same type of attracting set.
\end{abstract}

\maketitle

\noindent
{\bf Introduction.} \quad {\it Listeria monocytogenes} is a widely distributed 
pathogenic bacteria which occasionally causes serious illness in humans.  {\it 
L. monocytogenes} evades the host's immune system by living inside its cells.  
Proteins located on the surface of the rod shaped bacteria catalyze the 
polymerization of the infected cells' actin molecules and this activity propels 
the bacteria through the cytoplasm \cite{tilney}.  The underlying mechanism of 
actin-based motility is a subject of great interest both because {\it L. 
monocytogenes} is a deadly pathogen and because actin filament assembly plays a 
role in many forms of cell movement \cite{rafelski}.  A useful feature of 
actin-based motility in {\it L. monocytogenes} is the ``comet tail'' of actin 
filaments which are left behind as a cell is transported \cite{cameronb}.  The 
``comet tails'' provide a record of bacterial trajectories in the cytoplasm.  
These trajectories, which vary considerably between individuals, can be 
complicated and orderly at the same time.  In the present work it is shown how 
these trajectories can be explained with a dynamical system that has a well 
known type of attracting set.  The presence of this attracting set can account 
for the renewal of actin-based motility after the bacterium divides and the 
persistence of motility as the bacterium invades neighboring host cells. 

   {\it L. monocytogenes} trajectories are the result of a complicated
interaction of proteins in a host cell's cytoplasm or in a cytoplasmic extract
\cite{loisel}.  There has been extensive research into the molecular mechanism
underlying actin-based motility but they do not account for the complicated
forms of {\it L. monocytogenes'} actin comet tails \cite{peskin,noireaux,
gerbal,dickinson,mogilner,alberts,zeile,dickinsonb}.  In \cite{shenoy} Shenoy
{\it et al.} present a simple and remarkably effective kinematic model for the
trajectories of {\it L. monocytogenes} in a thin layer of cytoplasmic extract.

\begin{figure}
\includegraphics[width=8.7cm]{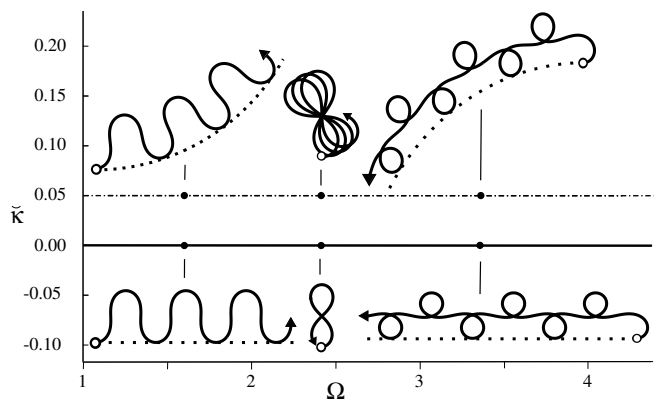}
\caption{Six curves in the $(x,y)$-plane determined by $d\theta/ds=
\breve{\kappa}+ \Omega\cos(s)$.  The inset for each $(x,y)$ curve points to its
corresponding parameter values $(\Omega,\breve{\kappa})$.  Each $(x,y)$ curve
starts at the point $(0,0)$ (marked by an open circle) in the direction
$\theta = 0$.  In two cases with $\breve{\kappa}=0$ the curves exhibit
linear drift (which follows the dotted lines).  The paths are qualitatively the
same for $\breve{\kappa}=1/20$ but show an overall tendency to veer from a
straight course (as indicated by the dotted curves).  The region of parameter 
space shown here only contains a portion of the parameter values that can 
replicate {\it L. monocytogenes} trajectories.} 
\label{fig.1}
\end{figure}

   In the Shenoy {\it et al.} model the effect of actin polymerization is 
assumed to produce a net force on the cell body which points slightly off 
center and which causes the bacteria to spin about its long axis as it travels 
in two dimensions.  The bacterium's velocity is given by a vector whose 
direction varies sinusoidally with time and whose magnitude is fixed.  Choosing
units of measure so the speed is 1, letting $s$ stand for arc length, and 
$\theta$ stand for the velocity's direction the Shenoy {\it et al.} model, in a 
non-dimensionalized form, is $d\theta/ds = \Omega \cos(s)$ where $\Omega \geq 
0$ represents the maximum deflection from forward motion.

     Since $d\theta/ds$ equals curvature the non-dimensionalized form of the
Shenoy {\it et al.} model gives a one parameter family of intrinsic equations
for planar curves ({\it i.e.} a two dimensional analog for the Frenet-Serret
equations) which exhibit qualitative changes as the parameter $\Omega$
is varied.  Shenoy {\it et al.} show that for small $\Omega$ the curve is
sinusoidal, for $\Omega\approx 2.5$ it resembles a figure eight, and for
larger values of $\Omega$ the curve tends to turn successively clockwise and
counter-clockwise around a sequence of points (fig.~\ref{fig.1}).  There are 
many qualitatively different types of curves for values of $\Omega$ from 0 to 
16 and Shenoy {\it et al.} show that {\it L. monocytogenes} display most if not
all of these types.

    Although this kinematic model can reproduce many of the complicated {\it L. 
monocytogenes} trajectories it does not explain how the forces on the cell body
arise or address the stability of the motion.  The question of how actin-based
motility arises is important because the bacteria have to divide in order to 
proliferate inside the host organism and after cell division the protein 
catalyzing actin polymerization is redistributed on the bacterial surface 
\cite{kocks,rafelski06}.  The issue of stability is important because one of 
the functions served by actin-based motility is to enable {\it L. 
monocytogenes} to create ``pseudopodal projections'' from one host cell into 
another and thus allow the bacteria to avoid the host's immune system 
\cite{tilney}.  If the dynamics underlying actin-based motility was not stable 
then obstacles in the bacteria's path could disrupt the bacteria's entry into 
the neighboring host cell.
  
   In the next section the model of Shenoy {\it et al.} is extended to allow
the angular displacement of {\it L. monocytogenes} trajectories to accumulate.
The third section explains how this extension leads to a dynamical system with
the same type of attracting set as seen with meandering spiral waves.  The
fourth section shows how the existence of the attracting set accounts for
actin-based motility developing and persisting in {\it L. monocytogenes}.

\ \\
\noindent 
{\bf 2.~Generalizing the kinematic model.} \quad It is well known in ballistics 
that it is difficult for a projectile to travel in a perfectly straight 
direction.  For a self propelled bacterium in a viscous medium even a small 
asymmetry in the cell body can cause it to eventually deviate from a straight 
course \cite{rafelskib}.  On the other hand in the Shenoy {\it et al.} model 
the angular displacement remains within fixed bounds along the entire length of 
the curve and the overall trajectory conforms to a straight line despite the 
small scale oscillations in its direction.  

   To improve the accuracy of the kinematic model it is worth considering
previous studies on actin-based motility.  Rutenberg and Grant \cite{rutenberg}
related the overall curvature of the trajectories to the number of randomly 
located actin filaments propelling the cell.  They treated the torque produced
by the filaments as a constant for relatively long periods of time which led to
trajectories with constant curvature, $d\theta/ds = \breve{\kappa}$.  For 
trajectories with $\breve{\kappa} \neq 0$ starting in the direction 
$\theta_0=0$ the angular displacement grows in proportion to arc length.   

   Evidence for the secular dependence of angular displacement on arc length in 
the Rutenberg and Grant model was subsequently presented in the extensive 
experimental study on actin-based motility in {\it L. monocytogenes} 
\cite{soo}.  The study found bacteria trajectories that were nearly circular 
and whose angular displacements and path lengths were nearly proportional to 
the elapsed time.  Consequently the angular displacements were nearly 
proportional to arc length. 

   Here we combine the approaches of Shenoy {\it et al.} and Rutenberg and 
Grant into a single model and present a summary of the types of paths this 
model displays.  The Shenoy-Rutenberg model is, in non-dimensional form, 
$d\theta/ds = \breve{\kappa}+ \Omega\cos(s)$ where $\Omega$, $\breve{\kappa}$ 
are constants.  For small $\breve{\kappa}$ the paths are qualitatively the same 
as for $\breve{\kappa}=0$ but they veer from a straight course 
(fig.~\ref{fig.1}).  Shenoy {\it et al.} found it useful to modify their model 
in a few cases by adding a low frequency term but this still did not allow the 
angular displacement to accumulate.  Adding a constant term as we do here does 
allow the angular displacement to accumulate and it helps shed light on the 
effectiveness of the Shenoy {\it et al.} model.

   The Shenoy-Rutenberg model is comparable in form to a model by Friedrich and
J\"{u}licher for the chemotaxis of sperm cells \cite{friedrich}.   Both models
determine the curvature of the path followed by the cells using a constant
curvature term and a second term but they differ in the form of the second
term.  For the Friedrich and J\"{u}licher model the second term is a function
of the chemoattractant concentration and the internal signaling network.  The
paths produced by the Friedrich and J\"{u}licher model depend on the form of
the concentration field.

      It is not very difficult to determine the paths produced by the 
Shenoy-Rutenberg model.  Let $(x(s),y(s))$ denote the arc length 
parameterization of a path.  The addition of $\breve{\kappa}$ leaves the model 
in the form of an intrinsic equation for planar curves so, for fixed values of 
the parameters, the solutions are congruent and we can focus on the initial 
condition $(x,y,\theta) = (0,0,0)$.  This gives
\begin{equation}
\label{eq.1}
\begin{pmatrix} 
x(s) \cr y(s)
\end{pmatrix} 
= \int_0^s
\begin{pmatrix}
\cos(\breve{\kappa} \sigma + \Omega \, \sin(\sigma)) \cr  
\sin(\breve{\kappa} \sigma + \Omega \, \sin(\sigma))
\end{pmatrix}
 d\sigma
\end{equation}
Changing the sign of either $\Omega$ or $\breve{\kappa}$ yields congruent
$(x,y)$ curves so we can assume $\Omega, \breve{\kappa} \geq 0$.  While the
integral cannot be evaluated in terms of elementary functions the curves are
symmetrical and made up of congruent copies of an arc of length $\pi$.  Since
the $(x,y)$ curve is invariant under reflection about the $y$-axis we can
reflect the arc for $0 \leq s \leq \pi$ to obtain the arc for $-\pi \leq s
\leq \pi$.

    For non-integral $\breve{\kappa}$ let $r = \cot(\pi \breve{\kappa}) x(\pi) 
+ y(\pi)$.  It can be shown that
\begin{equation}
\begin{small}
\label{eq.2}
\begin{pmatrix}
 x(s+2\pi) \cr y(s+2\pi) - r 
\end{pmatrix}
 =
\begin{pmatrix}
 \cos(2\pi \breve{\kappa}) &  -\sin(2\pi \breve{\kappa}) \cr   
          \sin(2\pi \breve{\kappa}) & ~~\cos(2\pi \breve{\kappa}) 
\end{pmatrix}
\begin{pmatrix}
 x(s) \cr y(s) - r
\end{pmatrix}
\end{small}
\end{equation}
From this it follows that the $(x,y)$ curve can be obtained by iteratively
rotating the arc for $-\pi \leq s \leq \pi$ about the point $(0,r)$ which is 
the center of symmetry for the figure.

  For non-integral rational $\breve{\kappa} = p/q \; (p,q$ coprime) and $\Omega
\neq 0$ the $(x,y)$ curve is closed with $q$-fold rotational symmetry.  It is
the union of $2q$ congruent arcs of length $\pi$.  For irrational
$\breve{\kappa}$ and $\Omega \neq 0$ the $(x,y)$ curve is quasiperiodic in the
plane.  It is the union of an infinite number congruent arcs with length $\pi$.

   For integer values of  $\breve{\kappa}$ we can think of $r$ as having gone
to infinity.  It can be shown that $y(s+2\pi)= y(s)$.  To express the value of
$x$ at multiples of $\pi$ we can use the integral representation for Bessel
functions
\begin{equation}
\label{eq.3}
   J_{\breve{\kappa}}(-\Omega) = \frac{1}{\pi}                     
\int_0^{\pi} \cos(\breve{\kappa} \sigma + \Omega \, \sin(\sigma)) \, d\sigma
\end{equation}
(this integral representation does not apply to non-integer values of
$\breve{\kappa}$).  From this it follows that the $(x,y)$ curve can be obtained
by iteratively translating the arc for $-\pi \leq s \leq \pi$ horizontally by
the distance $2\pi J_ {\breve{\kappa}}(-\Omega)$.  When $-\Omega$ is a zero of
$J_{\breve{\kappa}}$ the $(x,y)$ curve is closed with length $2\pi$.  Otherwise
the $(x,y)$ curve is the union of an infinite sequence of congruent arcs with
length $\pi$.  This generalizes a similar result from \cite{shenoy} for the
$\breve{\kappa}=0$ case.

   The points of maximal curvature on an $(x,y)$ curve occur where $s$ is an
even multiple of $\pi$, the points of minimal curvature occur where $s$ is an
odd multiple of $\pi$, and the curvature varies monotonically in between.  In a
neighborhood of $(0,0)$ an arc of the $(x,y)$ curve lies above the horizontal
tangent at $(0,0)$ and for $\breve{\kappa} > \Omega$ the curvature is positive
everywhere.

\begin{figure}
\includegraphics[width=8.7cm]{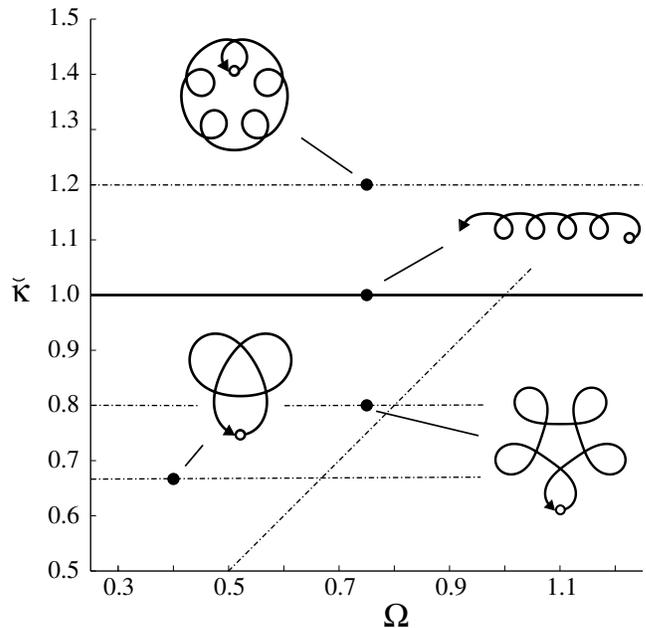}
\caption{A version of a Zykov-Winfree flower garden \cite{zykov,winfreeb} whose
isogonal contours have been combed straight.  The isogonal contours of 
$d\theta/ds= \breve{\kappa}+ \Omega\cos(s)$ for $\breve{\kappa} = 2/3,4/5,1,
6/5$ are shown.  Each $(x,y)$ curve in the insets starts at the point $(0,0)$ 
(marked by an open circle) in the direction $\theta = 0$.  For $\breve{\kappa} 
= 1$ the $(x,y)$ curves exhibit linear drift.  For $\breve{\kappa}$ below $1$ 
the $(x,y)$ curves have hypotrochoid like shapes and for $\breve{\kappa}$ above 
$1$ the $(x,y)$ curves have epitrochoid like shapes.  So long as 
$\breve{\kappa} > \Omega$ (above the diagonal line) the $(x,y)$ curves do not 
have inflection points.}
\label{fig.2}
\end{figure}

   For $1 = \breve{\kappa} > \Omega$ the $(x,y)$ curve has the form of a
trochoid with its ``petals'' lying in a row (fig.~\ref{fig.2}).  For
small $\Omega$ and $0 < \breve{\kappa} < 1$ the $(x,y)$ curve has the form of a
hypotrochoid with its ``petals'' on the outside.  For small $\Omega$ and $1 <
\breve{\kappa} < 2$ the $(x,y)$ curve has the form of an epitrochoid with its
``petals'' on the inside.

   Flower like curves such as these are traced out by the tips of spiral waves
propagating through excitable media.  Spiral waves occur in diverse systems
with very different underlying mechanisms.  This includes aggregating
myxobacteria which form macroscopic waves as cells glide across a two
dimensional surface \cite{reichenbach}.  A spiral wave often propagates as
though it were a rigid body rotating about a quiescent core.  Away from the
core the shape of the wave front converges to an Archimedean spiral
\cite{weiner,keener}.  However under appropriate circumstances the inner tip
undergoes a secondary oscillation as the wave rotates and thereby traces a
hypo/epi/trochoid like curve.  Spiral tip meander has been observed in many
systems such as the BZ chemical reaction \cite{winfreea}, heart tissue
\cite{ikeda}, and aggregating cells of {\it Dictyostelium discoideum} (cellular
slime molds) \cite{foerster1990}.

\ \\
\noindent
{\bf 3.~Resonant Hopf bifurcations.} \quad An important step toward 
understanding why spiral tip meander occurs in systems with such different 
underlying mechanisms was made by Barkley \cite{barkley1992,barkley1994,
barkley1994b} who recognized, through numerical and mathematical analysis, the 
important role played by the group of orientation preserving congruences of the 
Euclidean plane, ${\bf SE}(2)$, and that this role can be exemplified by 
reducing the dynamics to five dimensions.  The mathematics of Barkley's 
breakthrough has been further elaborated and generalized \cite{wulff,fiedler,
sandstede,golub,golubb}.  Barkley's approach can be nicely illustrated with the 
Shenoy-Rutenberg model since it already has the form of an intrinsic equation 
for planar curves.  To do this we couple the Shenoy-Rutenberg model to a two 
dimensional system from \cite{holmes}.  We write the Cartesian coordinates for 
this subsystem as $(X,Y)$.  The full differential equation is 
\begin{eqnarray} 
\label{eq.4}   
    x' &=& \cos(\theta) \nonumber \\
    y' &=& \sin(\theta) \nonumber \\         
    \theta' &=& \breve{\kappa} + X \\
    X' &=&   -Y + (\mu - X^2-Y^2)X \nonumber \\
    Y' &=& \,\,\,\, X + (\mu - X^2-Y^2)Y \nonumber
\end{eqnarray}
where the parameter $\mu$ corresponds to Barkley's normalized bifurcation 
parameter \cite{barkley1994}.  For $\mu<0$ the origin of the $(X,Y)$ subsystem 
is an attracting fixed point.  At $\mu=0$ the Hopf bifurcation occurs and for 
$\mu>0$ there is an attracting circular limit cycle centered at the origin with 
radius $\sqrt{\mu}$.  

   To concisely express the initial conditions we set $\Omega = 0$ for $\mu < 
0$ and $\Omega = \sqrt{\mu}$ for $\mu \geq 0$.  For the initial condition 
$(x,y, \theta,X,Y)= (0,0,0,\Omega,0)$ the solution to the $(X,Y)$ subsystem is 
$(X(s),Y(s))= \Omega \, (\cos(s),\, \sin(s))$ which gives $\theta' = 
\breve{\kappa} + \Omega \cos(s)$ which in turn recovers eq.~(\ref{eq.1}) for 
the $(x,y)$ subsystem.

   A purely rotating spiral wave appears motionless in a frame rotating with
it.  The transition to meandering corresponds to the Hopf bifurcation.  After
the bifurcation the spiral tip appears in the rotating frame to trace a
circularly shaped path although far from the core the wave continues to appear
motionless.

   By converting eq.~(\ref{eq.4}) to a rotating coordinate system $(0,0,0,
0,0)$ becomes a fixed point with spectrum $\{\pm i\breve{\kappa}, \, 0,
\, \mu\pm i\}$.  The eigenvalues $\pm i\breve{\kappa}$ arise from the
translational symmetry of the plane and $0$ arises from the rotational symmetry
of the plane.  At the Hopf bifurcation all five eigenvalues lie on the
imaginary axis.

     Barkley showed that the type of curve traced by a spiral tip in the
stationary frame depends on where the Hopf eigenvalues cross the imaginary axis
in relation to the translational eigenvalues.  When the translational
eigenvalues are between the Hopf eigenvalues the spiral tip will follow a
hypotrochoid like curve ($0< \breve{\kappa} < 1$ in eq.~(\ref{eq.4})).
When the translational eigenvalues are outside of the Hopf eigenvalues (but
not more than twice the Hopf eigenvalues) the spiral tip will follow an
epitrochoid like curve ($1< \breve{\kappa} < 2$ in eq.~(\ref{eq.4})).  When
the translational and Hopf eigenvalues coincide the spiral tip exhibits linear
drift ($\breve{\kappa} = 1$ in eq.~(\ref{eq.4})).

   In terms of {\it L. monocytogenes} we can interpret $(X,Y)$ as the
projection of the cell's translational velocity to a plane orthogonal to the
cell body's long axis and we can interpret the oscillation of $(X,Y)$ as the 
effect of the cell's spin on its propulsion system.  The long axis and the
$X$ component are parallel to the surface being traversed while the $Y$
component points in the orthogonal direction.  For a cell constrained in two
dimensions the $Y$ component does not contribute to the motion.  For $\Omega=0$
the cell appears motionless in a frame rotating with it.  For small $\Omega>0$
the cell appears to follow a circularly shaped path in the rotating frame.

 The detailed mechanisms behind spiral meander and {\it L. monocytogenes}
motility are different but the paths they follow are both part of a larger two
parameter family of curves.  The paths followed by spiral wave tips are
organized around a first order resonant Hopf bifurcation for which the
translational and Hopf eigenvalues coincide ($\breve{\kappa} = 1$ in 
eq.~(\ref{eq.4})).  The paths followed by {\it L. monocytogenes} are organized
around a zero order resonant Hopf bifurcation for which the translational and
rotational eigenvalues coincide ($\breve{\kappa} = 0$ in eq.~(\ref{eq.4})).

\ \\
{\bf 4.~The dynamics of {\it L.} monocytogenes motility.}  \quad Spiral waves 
appear in excitable media when, in the state space for the medium, the state is 
sufficiently close to the appropriate attracting set.  Each state in the 
attracting set corresponds to a well formed spiral wave in the medium.  In many 
cases there is a characteristic wavelength to the limiting form of the spiral 
wave \cite{winfree72}.  In such cases any two spiral waves in a planar 
homogenous isotropic medium will be congruent.  For media which support 
non-meandering spiral waves the attracting set is essentially a copy of the 
symmetry group ${\bf SE}(2)$.  This is the type of attracting set the solutions 
to eq.~(\ref{eq.4}) have when $\mu<0$.  The orbits of the dynamical system 
inside the attracting set are simple closed curves and thus bounded.  These 
orbits correspond to spiral waves undergoing a pure rotation or {\it L. 
monocytogenes} following a circular trajectory.   

    Aside from numerical simulations it is difficult to prepare excitable media
so that the initial state of the system is within the attracting set, 
{\it i.e.} so that the medium begins with a well formed spiral wave which then
undergoes a pure rotation.  A purely rotating spiral wave only appears after a
transient period.  One way for a purely rotating spiral wave to appear is by
disrupting a circular or linear wave front with an obstacle in the medium.  The
broken end of the wave front will then curl up and over time the shape of the
wave will develop into a well formed spiral which propagates in a purely
rotational manner.  The disruption of a wave front by an obstacle brings the
state of the system sufficiently close to the attracting set that it converges
towards it.

    For homogenous isotropic excitable media which support meandering spiral 
waves the attracting set has another dimension.  This is the type of attracting 
set the solutions to eq.~(\ref{eq.4}) have when $\mu>0$.  Each point in the 
attracting set gives the position and orientation of the spiral wave as well as 
its phase within the period of meander.  The orbits of the dynamical system 
inside the attracting set are bounded unless there is a resonance between the 
rotation of the spiral wave and the oscillation of the tip in which case they 
are unbounded.  Meandering spiral waves appear after a transient period once 
the state of the system has been brought sufficiently close to the attracting 
set.  For systems at or near resonance the core of the spiral wave will be 
transported across large distances.

   Biological systems repeat many of the same developmental strategies in 
various contexts to form functional patterns.  The presence of low dimensional 
attracting sets in complicated dynamical systems can provide stability to 
developmental processes which are exposed to the environment.  For instance 
there are prokaryotes ({\it e.g.} myxobacteria) and eukaryotes ({\it e.g. D. 
discoideum}) which use spiral wave dynamics to get individual cells dispersed 
over a wide area to aggregate together and develop multicellular reproductive 
organs.  The underlying mechanisms by which myxobacteria and {\it D. 
discoideum} move and communicate are quite different but the presence of an 
attracting set for spiral wave dynamics can allow the aggregation process to 
proceed despite the vagaries of their environments.

  There has been a long running and continuing effort to determine the 
underlying mechanism responsible for the formation of actin-based motility
\cite{hill,peskin,noireaux,gerbal,dickinson,mogilner,alberts,zeile,dickinsonb}.
The Shenoy {\it et al.} model is effective at duplicating {\it L. 
monocytogenes} trajectories but it is not directly based on a physicochemical 
mechanism.  Their model proceeds from general considerations about how the 
forces produced by actin polymerization must act on the cell.  In order for the 
cell to change direction as it moves there must be some asymmetry in the 
distribution of forces exerted on the cell surface.  By treating the net 
propulsive force as a constant parallel to the long axis of the cell and whose 
exertion point rotates at a constant distance about the long axis the magnitude 
of the component of the net torque orthogonal to the plane of motion varies in 
a precisely sinusoidal fashion.  In this way the cell body oscillates about its 
center of mass much like an ideal torsional spring.  With the propulsive force 
always parallel to the long axis the cell moves in trajectories that 
alternately wind clockwise and counter-clockwise.  

   However this clockwork like mechanism does not simply appear fully formed. 
For {\it L. monocytogenes} engaged in actin-based motility the concentration of 
ActA (the catalyst for actin polymerization on {\it L. monocytogenes}) along 
the cell wall increases from the apical pole to the basal pole.  {\it L. 
monocytogenes} cells reproduce by dividing along a septum midway between the 
poles.  After division each daughter cell forms a new apical pole at the 
septation region and ActA concentration is redistributed along the cell wall 
\cite{kocks,rafelski06}.

   When a {\it L. monocytogenes} cell first begins moving it engages in a 
``hopping'' type motion.  The density of actin builds up behind the basal pole 
until there is a sudden acceleration of the cell.  The actin tail then becomes 
rarefied, the cell subsequently decelerates nearly to rest, and the cycle 
repeats.  After several cycles the cell eventually settles down to a relatively
constant speed \cite{gerbal,rafelskib}.

   One function served by the actin-based motility of {\it L. monocytogenes} is 
the transport of bacteria from one host cell to another without the bacteria 
having to leave the confines of the host cells.  This is accomplished when a 
bacterium presses against the host's plasma membrane to create a ``pseudopodal 
projection'' with the bacterium inside.  The bacterium enters a neighboring 
host cell when the pseudopodal projection is phagocytosed by the neighboring
cell \cite{tilney}.
   
   During the transient period {\it L. monocytogenes} movement is sensitive to 
obstacles in its environment but it becomes more robust when it reaches a 
steady speed \cite{rafelskib}.  The presence of a low dimensional attracting 
set for actin-based motility can provide stability in the development of {\it 
L. monocytogenes} infectiousness.  When the state of the system is in its 
transient phase, away from the attracting set, obstacles in the environment can 
have the effect of perturbing the system from one orbit to another with a very 
different course.  On the other hand when the state of the system is close to 
the attracting set it can quickly return to the attracting set after an 
obstacle causes a perturbation.  This accounts for how {\it L. monocytogenes} 
motility can persist as it forms a pseudopodal projection.

   Moreover the attracting set for actin-based motility by {\it L. 
monocytogenes} confined to two dimensions appears to be of the same type as for 
spiral tip meander in two dimensions.  In both cases the points of the 
attractor correspond to the position, the orientation, and the phase in the
secondary oscillation of the object that is moving.  For the purposes of 
transport it is useful for the system to be near a resonance.  The Shenoy {\it 
et al.} model corresponds to a zero order resonance which helps account for its
effectiveness although the model needs to be placed in a larger mathematical
context to see this and to account for how actin-based motility in {\it L. 
monocytogenes} arises and persists in the presence of obstacles.

\ \\
{\bf Conclusion.} \quad Spiral waves involve the cooperative behavior of 
multiple agents while a bacterium is generally thought of as a single agent.  
However this apparent association of the first order resonant Hopf bifurcation 
to the organized motion of multiple agents and the zero order resonant Hopf
bifurcation to the motion of a single agent is not a general principal.  The 
movement of an individual bacterium can be regarded as the action of a single 
agent but actin-based motility involves the polymerization of actin so it can 
also be regarded as an organized activity involving many chemical agents.  
Inert beads coated with proteins which catalyze actin polymerization can also 
engage in actin-based motility \cite{cameron,shaevitz}.

   Actin is an important constituent of the cytoskeletons of eukaryotic cells 
and it forms locomotory structures such as filopodia in which actin organizes 
into bundles and lamellipodia in which actin organizes into meshworks 
\cite{heath,small1995,anderson,svitkina,small2002,dent,pollard}.  Numerical 
simulations indicate that actin can also form spiral waves \cite{karsten} and 
there is evidence for actin forming spiral waves inside of {\it D. discoideum} 
pseudopodia \cite{vicker,whitelam}.  
  
   The actin rich cytoplasm of eukaryotic cells can be seen as a type of 
excitable media with a propensity to locomotory behavior which several 
pathogens, in addition to {\it L.  monocytogenes}, have evolved to take 
advantage of in order to transport themselves.  Actin-based motility occurs 
with the bacteria {\it Shigella flexneri} \cite{bernardini}, species 
of {\it Rickettsia} bacteria \cite{heinzen}, and vaccinia virus particles 
\cite{cudmore}.  The detailed mechanism of actin-based motility varies between 
these pathogens but they each involve catalyzing the polymerization of actin, 
\cite{kocks1995,vanKirk, gouin}.

    When {\it L. monocytogenes} bacteria are constrained to move in two 
dimensions their trajectories can form a complicated series of coils.  As with 
the trajectories of meandering spiral waves, the trajectories of {\it L. 
monocytogenes} are composed of essentially congruent copies of a finite curve 
repeatedly joined end to end.  The form of the repeating unit appears after a 
transient period and it varies between different occurrences of meandering 
spiral waves and between different individual {\it L. monocytogenes}.  Yet 
complicated curves characteristic of both types of phenomena can be replicated 
using a two parameter family of dynamical systems with the same type of 
attracting set as eq.~(\ref{eq.4}).  Taken together these lines of evidence 
support the idea that actin-based motility in {\it L. monocytogenes} is 
organized by the same type of low dimensional attracting set that organizes 
spiral tip meander.

\ \\
\noindent
{\bf Acknowledgment.} I would like to thank Omar Clay and Jeff Yoshimi for 
their comments and suggestions to improve the presentation.

\ \\
\noindent
scotton@sdf.lonestar.org

\end{document}